\documentclass[prd,  showpacs, eqsecnum, floatfix, superscriptaddress, twocolumn, longbibliography, nofootinbib]{revtex4-2}

\usepackage{bm, amssymb, amsmath, siunitx, color, booktabs } 
\usepackage[mathscr]{euscript}

\usepackage[english]{babel}
\usepackage[T1]{fontenc}
\usepackage{graphicx}
\graphicspath{{./figures/}}

\DeclareSIUnit[]\solarmass{\text{\ensuremath{{\textup{M}}_\odot}}}
\DeclareSIUnit[]\year{yr}

\usepackage{hyperref}
\hypersetup{
	colorlinks=true, 	% false: boxed links; true: colored links
	linkcolor=cyan,		% color of internal links
    citecolor=blue,		% color of links to bibliography
}

\newcommand{\nth}{\textsuperscript{th}}

\newcommand{\norm}[1]{\left\lVert#1\right\rVert}

\newcommand\innerprod[2]{\langle #1 \,|\, #2 \rangle}

\def \fFactor {\mathcal{F\!F}}

\def \msun	{{M}_\odot}

\def \flow	{f_{\mathrm{low}}}

\def \fhigh	{f_{\mathrm{high}}}
\def \dmax {{\mathcal{D}_{\mathrm{max}}}}
\def \mmin {\mathcal{MM}}

\def \pycbc {\textsc{PyCBC}}
\def \iFAR {\textsc{iFAR}}

\newcommand{\anstar}[1]{\mathscr{A}_{#1}^{\ast}}

\newcommand{\IITGn}{Indian Institute of Technology Gandhinagar, Gujarat 382355, India.\vspace*{0.135cm}}
\newcommand{\NIKHEF}{Nikhef, Science Park 105, 1098 XG Amsterdam, The Netherlands.}
\newcommand{\UU}{Institute for Gravitational and Subatomic Physics (GRASP), \mbox{Utrecht University}, Princetonplein 1, 3584 CC Utrecht, The Netherlands.}

\begin{document}

\title{Beyond general relativity: designing a template-based search \mbox{for exotic gravitational wave signals}}

\author{Harsh~Narola} \email[e-mail: ]{h.b.narola@uu.nl}

%\affiliation{\IISERTirupati}
\affiliation{\IITGn}
\affiliation{\NIKHEF}
\affiliation{\UU}
\author{Soumen~Roy} 
\affiliation{\IITGn}
\affiliation{\NIKHEF}
\affiliation{\UU}
\author{Anand~S.~Sengupta\vspace*{0.25cm}} 
\affiliation{\IITGn}

\begin{abstract}
%Detecting gravitational waves from compact binary sources has stimulated the study of extreme astrophysical phenomena and exploring fundamental physics in the nonlinear strong-field regime of gravity hitherto inaccessible to previous experiments. 
%
Accurate waveform models describing the complete evolution of compact binaries are crucial for the maximum likelihood detection framework, testing the predictions of General Relativity (GR) and investigating the possibility of an alternative theory of gravity.
Deviations from GR could manifest in subtle variations of the numerical value of the GW signal's post-Newtonian (PN) phasing coefficients. Once the search pipelines confirm an unambiguous signal detection, deviations of the signal phasing coefficients at various PN orders are routinely measured and reported. 
As the search templates themselves do not incorporate any deviations from GR, they may miss astrophysical signals carrying a significant departure from general relativity. 
We present a parametrized template-based search for exotic gravitational-wave signals beyond General Relativity by incorporating deviations to the signal's phasing coefficients at different post-Newtonian orders in the search templates. We present critical aspects of the new search, such as improvements in search volume and its effect on various parts of the parameter space. In particular, we demonstrate a factor x2 increase in search sensitivity (at a fixed false-alarm rate) to beyond-GR exotic signals by using search templates that admit a range of departures from general relativity.
We also present the results from a re-analysis of the 10-days long duration of LIGO's O1 data, including the epoch of the GW150914 event, highlighting the differences from a standard search. We indicate several directions for future research, including ways of making the proposed new search computationally more efficient.

\end{abstract}
\maketitle

%%%==== SECTION ====%%%
\section{Introduction}
\label{sec:intro}

%-----------
The direct detection of gravitational waves (GWs) from the merging compact binary systems of black holes and neutron stars have ushered in a new era of gravitational-wave astronomy~\cite{gw150914.PhysRevLett.116.061102, LIGOScientific:2016dsl, GWTC1-PhysRevX.9.031040, GWTC-2-Abbott:2020niy}. While the direct detection of GW itself can be considered to be an independent test of General Relativity (GR),  such signals also provide a unique opportunity to test fundamental physics in the highly dynamical, nonlinear strong-field regime of gravity inaccessible to the previous, successful experimental tests of GR such as the perihelion precession of Mercury's orbit~\cite{MTW}, deflection of light by a massive object~\cite{lensing}, and the gravitational redshift of light~\cite{redshift}. None of the previous tests could probe GR in extreme environments as provided %by observing signals from the final inspiral and the final merger 
by the signals from the merger of two compact objects.
%GW detections also open up a unique way to shed light on the dark side of our Universe. 
Like the previous experiments, tests of GR performed using GW signals \cite{TGR-gwtc-1} strongly support the predictions made by GR. The success of these tests has established GR to be the most accurate theory of gravity at present, leading us to speculate that the correct theory of gravity may not be too far from GR. 
 
Gravitational wave detection methods can be divided into two broad categories: template-based searches and generic transient searches. The template-based searches rely on accurate theoretical models of the signal waveform, a great scientific challenge for several decades. Recent developments in high-order post-Newtonian (PN) calculations~\cite{Blanchet:2013haa}, quasi-normal modes in the ringdown phase~\cite{Berti:2005ys}, and numerical relativity (NR)~\cite{Pretorius:2005gq, Campanelli:2005dd, Baker:2005vv} simulations have made it possible to generate accurate semi-analytical waveforms that describe the dynamics of inspiral, merger, and subsequent ringdown of the remnant object~\cite{Buonanno:1998gg, Buonanno:2009qa, Pan:2011gk, Taracchini:2012ig, Taracchini:2013rva, Ajith:2007qp, Ajith:2007kx, Santamaria:2010yb, Ajith:2009bn, Khan:2015jqa, Husa:2015iqa}. These studies focus on modelling %astrophysically relevant 
compact binaries consisting of black holes and neutron stars in the theory of GR. These are commonly used as template waveforms in standard GW search pipelines~\pycbc\cite{Usman_2016}. These search pipelines are carefully tuned to identify signals that match the template waveforms embedded in noisy data recorded by GW detectors \cite{KAGRA:2013rdx} while efficiently rejecting noise artefacts that mimic GW signals.

Accurate template waveforms is a crucial input to template-based searches as inaccuracies in modelling the dynamics of the two-body problem can lead to a severe loss in detection efficiency. While it may still be possible to detect considerably strong transient GW signals using the techniques of `burst searches' that are agnostic to signal models, accurate waveform models are still required to infer source parameters (such as component masses, spins and distance to the source from Earth, etc.) and eliminate any bias in these measurements~\cite{Cornish:2011ys}. %Therefore, it is desirable to have a sound knowledge of GW signal models in beyond-GR theories: not only to improve gravitational wave astronomy but also to find out a better theory of gravity.
Several promising alternative theories to GR have been developed in the last century, such as the scalar-tensor, Brans-Dicke and Chern-Simons theories all of which have survived a wide range of experimental tests. However, in nearly all alternative theories of gravity, one lacks a detailed understanding of the dynamics of the coalescing compact binary system in the strong-field regime. Recent advances in analytical approaches and NR simulations have made it possible to model binary black hole mergers and propagation of the GWs in the ``beyond-GR'' theories~\cite{Okounkova:2017yby, Witek:2018dmd, Okounkova:2019dfo}. However, it is not clear whether the full theories are well-posed. Also, there might be a more accurate alternative theory unknown to us to date.

Impelled by the lack of concrete knowledge of beyond-GR theories, we generally perform tests that allow a broad class of possible departure from GR by introducing parametrized deviations from the basic GR waveform. For example, the parametrized post-Newtonian (pPN) formalism allows deviations in the coefficients at different orders of velocity parameter ($v/c$) in the PN series of phase expansion. This is motivated by the fact that accurate reconstruction of the signal's phase evolution would be the most sensitive measurement for capturing any departure from GR.
% since the waveform being phase throughout the inspiral is salient for achieving a high fitting-factor. 
%Several promising alternative theories of gravity exhibit the imprints of non-GR effects in the phase evolution of GW signals resulting from corrections to the conservative and dissipative quantities. 
The non-GR effects are imprinted on the phase evolution of GW signals in several promising alternative theories of gravity arising from corrections to the expressions for conservative and dissipative quantities. For example, the scalar-tensor theory allows for the $-1$~PN term as the leading dipolar contribution and $0.5$~PN term for the tail-induced dipole effect~\cite{Yagi:2015oca, Yagi:2016jml, Barausse:2016eii, Arun:2012hf}, massive gravity shows contributions to the phase at 1PN~\cite{Will:1997bb}.

In this paper, we develop the method for a bottom-up search for signals carrying deviations in the PN phasing coefficients from those predicted by GR. We refer to these deviated signals as \textit{non-GR} signals. Such an approach allows us to look for generic deviations from GR.
This strategy contrasts with the top-down approach, where one starts with a particular theory of gravity, writes down the action of the theory, solves the equations of motion, and predicts some \textit{observables} which may or may not agree with the experimental data. Whereas in the bottom-up approach, one starts by assuming GR as a null hypothesis, adds deviations in the observables predicted by GR, and tests them against the experimental data collected by gravitational wave detectors. % For the present case, the observables are the PN phasing coefficients.  
If the bottom-up approach detects a plausible deviation in the GR-predicted observables then one can narrow down the space of all possible modifications to the GR action that are consistent with such deviations. 

GW events detected by LIGO and VIRGO detectors have led to data-driven constraints on the deviation of PN coefficients \cite{ptest-1-PhysRevD.74.024006, ptest-2-PhysRevD.82.064010, ptest-3-PhysRevD.85.082003} from their GR-predicted values. One can measure the deviation in the phasing coefficients at each PN order independently using Bayesian inference methods. 
%These constraints  
%from %performing the parameter estimation of 
%measurements of  using 
%
The analysis of events in the GWTC-1 and GWTC-2 catalogues indicate that the null-deviation hypothesis (hence, GR) lies within the 90\% credible interval of the posterior distribution over possible excursions of the phasing coefficients. However the width of these distributions also indicate a non-negligible range of deviations (see Fig. \ref{fig:jointDist}) that are also consistent with the data. In this paper we consider a range of possible deviations (in the phasing coefficients) as inferred from the combined analysis of previously detected events (and in this sense, the allowed domain of "non-GR signals" considered in our analysis is not wholly arbitrary). One recovers the base GR signal when all the deviation parameters are set to zero. We'd like to stress that the deviations introduced in the signal's phase coefficients to mimic non-GR signals are %generic and thereby 
generic and independent of any specific alternative beyond-GR theory of gravity.

%To answer these questions, we present the fitting factor calculation and PyCBC runs with three weeks of O1 data for two different cases: GR template bank against non-GR injections and non-GR template bank against non-GR injections. Before this, we perform a sanity check to confirm that the GR template bank we are using is efficiently recovering GR signals.  

This paper is structured as follows: we start by summarising the template-based search pipeline in Section \ref{sec:pipline}. This is followed by a discussion of the non-GR waveform model in Section \ref{sec:PS and WM} where we also motivate the choice for the search parameter space. In Section \ref{sec:bank}, we present the details of the construction and validation of the GR and non-GR template banks and show via a fitting factor analysis that the latter adequately cover non-GR signals. Results from a Monte-Carlo study are presented in Section \ref{sec:O1 data runs}, where we present and highlight improvements to the distance reach of the search pipeline by using non-GR templates. We re-analyze 10-days of data from advanced-LIGO's O1  \cite{RICHABBOTT2021100658} science run and present the results in section \ref{sec:O1_analysis}. Finally, we present our concluding remarks in Section \ref{sec:disc} and sketch the broad directions of future investigations following the methods presented in this paper.

%%%==== SECTION ====%%%
\section{Review of template based GW searches}
\label{sec:pipline}

Detecting a gravitational wave signal buried in the noisy data is a complex task since the amplitude of the gravitational wave signal is comparable to the detector noise level. %This task is performed by different \textit{search pipelines}. 
Many search pipelines are being developed for the last three decades. The search methodology of these pipelines can be divided into two broad categories - template based search and generic transient search. PyCBC \cite{Usman_2016} is one such search pipeline which performs a template based search. %The purpose of a PyCBC search is to identify a gravitational wave candidate event and assign a measure of statistical significance to it.
The PyCBC framework is designed to identify a gravitational wave candidate event and assign a measure of statistical significance to it.
The basic assumption in template based search is that the gravitational wave signal buried under the detector noise is very well approximated by the template waveforms. A waveform is parametrized by the system's intrinsic properties: component masses and spins, and extrinsic properties: sky location, distance to the source, inclination angle (angle between the total angular momentum and the line-of-sight), polarization angle, time, and phase.  Since the parameters of an incoming GW signal are not known a priori, the PyCBC uses a collection of template waveforms, called a template bank, to perform the search. The template bank spans the desired parameter space. We explain the process of template bank construction in section \ref{sec:bank}. The parameter space is the space of intrinsic and extrinsic parameters over which we want to conduct the search. We discuss this in more detail in section \ref{sec:PS and WM}. The search pipeline calculates the correlation of each template in the template bank with the data to construct a signal-to-noise (SNR) ratio $\rho(t)$ time series. This process is called \textit{matched filtering}.
%We have to rely approximation methods because we do not have an exact theoretical waveform predicted by GR. It is possible to construct template waveforms using numerical relativity simulations. However, the computational cost of these simulations makes them unfeasible. We have to use \textit{hybrid} template waveforms which are evaluated by matching the post-Newtonian description of the inspiral to a set of numerical relativity simulations. IMRPhenom is one such family of phenomenological template waveforms \cite{phenomPhysRevD.77.104017}, which we are going to use for the present exercise. In the absence of the exact template waveforms, \IMRPhenomD templates are used as proxy, which are not only \textit{effectual} to detect the gravitational wave signals but also \textit{faithful} in estimating the parameters of the binary.% 
%\begin{equation}
%\label{equ:snr}
%    \rho^2(t) = \frac{ \langle s \mid h_{\cos} \rangle^2 }{ \langle h_{\cos} \mid h_{\cos} \rangle} + \frac{ \langle s \mid h_{\sin} \rangle^2}{\langle h_{\sin} \mid h_{\sin} \rangle }
%\end{equation}
%\begin{equation}
%\label{equ:snr}
%    \rho^2(t) = \frac{ \innerprod{ s }{\hp}^2  }{ \innerprod{\hp}{\hp} } + \frac{ \innerprod{ s }{\hc}^2  }{ \innerprod{\hc}{\hc} }
%\end{equation}
\begin{equation}
\label{equ:snr}
    \rho^2 \equiv  \innerprod{ s }{\hat{h}_+ }^2   + \innerprod{ s }{\hat{h}_\times}^2 
\end{equation}
where the quantities $\hat{h}_+$  and $\hat{h}_\times$ are the unit-norm, linearly independent `plus' and `cross' polarization of the template waveform $\hat{h}$. 
%They are linearly independent. The hat on $\hat{h}$ denotes the normalized waveform, for example for plus polarization it is $\hat{h}_+ = h_+/\innerprod{h_+}{h_+}^{1/2}$. %denotes the normalized waveform, $\hat{h}_{+, \times} = h_{+, \times}/\innerprod{ h_{+, \times} }{h_{+, \times}}^{1/2}$. %The inner product between two time-series vectors (with a relative circular time-shift $\Delta t$ between them) is defined by 
%
%\begin{equation}
%    \langle s \mid h \rangle (\Delta t) = 4 \, \operatorname{Re} \left [ \int\limits_{f_\text{low}}^{f_\text{high}} \frac{\tilde{s}(f)\, \tilde{h}^*(f)}{S_n(f)}
%    \, e^{2\pi if \Delta t} \, df \right ]
%\end{equation}
%\begin{equation}
%    \innerprod{s}{h} (\Delta t) = 4 \, \Re  \int_{\flow}^{\fhigh} \frac{\tilde{s}(f)\, \tilde{h}^\ast(f)}{S_n(f)}
%    \, e^{2\pi if \Delta t} \, df 
%\end{equation}
The inner product between two time-series vectors is defined by 
\begin{equation}
    \innerprod{a}{b}(t) = 4 \, \Re  \int_{\flow}^{\fhigh} \frac{\tilde{a}^\ast(f)\, \tilde{b}(f)}{S_n(f)} e^{-i2\pi f t}\, df 
\end{equation}

% \begin{equation}
 %   \innerprod{a}{b} = 4 \, \Re  \int_{\flow}^{\fhigh} \frac{\tilde{a}^\ast(f)\, \tilde{b}(f)}{S_n(f)} \, df 
%\end{equation}
%where $\tilde{s}(f)$ is the Fourier transformed detector data defined by 
%\begin{equation}
%\tilde{s}(f) = \int_{-\infty}^{\infty} s(t) \, e^{- 2\pi ift} \, dt,
%\end{equation}
%and $\tilde{h}(f)$ is the Fourier transformed template waveform.
where $\tilde{a}(f)$ and $\tilde{b}(f)$ are the Fourier transform of $a(t)$ and $b(t)$ respectively and $S_n(f)$ is the one-sided detector noise power spectral density (PSD).% obtained by: 
%
%\begin{equation}
%\mathbb{E} \left [ \tilde{s}(f) \, \tilde{s}(f') \right ] = \frac{S_n(f)}{2} \, \delta(f-f'),
%\end{equation}
%
%where $\mathbb{E}$ denotes the ensemble average over several noise realizations and $\delta$ is the Dirac delta function. It is assumed that the noise is wide-sense stationary. The two phases of the template waveform are related by $\tilde{h}_{+}(f) \propto i \, \tilde{h}_{\times}(f)$. The cutoff frequencies, namely, $\flow$ and $\fhigh$ depend on the detector's sensitive bandwidth.

After computing the SNR time-series $\rho(t)$ for each template in each detector, the pipeline yields a list of timestamps when the SNR exceeds a predetermined threshold, which are called triggers. We expect that a signal may be present at that time. The data taken from gravitational wave detectors is neither Gaussian nor stationary. The noise artifacts can produce a large SNR without having good match between the data and the filter waveform. One of the most common method to distinguish between the real astrophysical signal and noise artifacts is to check whether the morphology of the signal in the data is consistent with the filter waveform that produced a trigger, which is called signal-consistency test~\cite{Allen:2004gu, Allen:2005fk}. We obtain a $\chi^2$ value by dividing the filter waveform into many non-overlapping segments with equal power and evaluate the level of agreement in each segments:
\begin{equation}
\chi^2 = p \sum_{i=1}^{p} \norm{ \innerprod{ s }{\hat{h}_i } - \innerprod{ s }{\hat{h} }/ p }^2,
\end{equation}
where, the filter waveform $\hat{h}$ is divided into $p$ pieces such that $\sum_{i=1}^{p}\hat{h}_i = \hat{h}$. If we assume the detector output is characterized by Gaussian noise and the additional signal is well approximated by the filtered waveform $h$, then the above statistic follows a $\chi^2$ distribution with a degrees freedom of $2p-2$. On the other hand, non-Gaussian noise artifacts or a lower match between the signal and filter waveform would lead to a large value of that statistic~\cite{Allen:2004gu, Allen:2005fk}. Many different techniques are adopted for analyzing the LIGO-Virgo data. All of them combines the matched filter SNR and the $\chi^2$ value to produce a ranking statistic~\cite{Babak:2012zx, Adams:2015ulm, Nitz:2017svb, Messick:2017}. %We adopt the reweighted SNR statistic $\hat{\rho}$ for our analysis, which has been used for analyzing the Advanced LIGO-Virgo data by $\pycbc$~\cite{LIGOScientific:2016dsl},
The recent searches of the Advanced LIGO's O1 data by $\pycbc$ employed the reweighted SNR statistic~\cite{LIGOScientific:2016dsl},
\begin{equation}
\label{eq:reweighted}
\hat{\rho}= \begin{cases}\rho /\left[\left(1+\left(\chi_{r}^{2}\right)^{3}\right) / 2\right]^{1/6}, & \text { if } \chi_{r}^{2}>1 \\ \rho, & \text { if } \chi_{r}^{2} \leq 1\end{cases},
\end{equation}
where the quantity $\chi_r^2 = \chi^2/(2p-2)$. Subsequently, the $\pycbc$ pipeline used a sine-Gaussian $\chi^2$ discriminator to produce the ranking statistic for analyzing the Advanced LIGO-Virgo data from O2 and O3 runs~\cite{GWTC1-PhysRevX.9.031040, GWTC-2-Abbott:2020niy, LIGOScientific:2021djp, Nitz:2020oeq, Nitz:2021uxj}, introduced in~\cite{Nitz:2017lco}. The main focus of incorporating the sine-Gaussian feature is to distinguish many blip glitches from short duration gravitational-wave signals. We adopt the sine-Gaussian reweighted SNR statistic $\hat{\rho}_{\rm{sg}}$ for our analysis. Therefore, we accept those triggers that lie above a pre-decided threshold of $\hat{\rho}_{\rm{sg}}$.
%\begin{equation}
%\hat{\rho}_{\rm{sg}}= \begin{cases}\hat{\rho}/\sqrt{\chi_{r,\rm{sg}}^2 /4} , & \text { if } \chi_{r,\rm{sg}}^2\geq 4 \\ \rho, & \text { if } \chi_{r,\rm{sg}}^2\leq 4\end{cases},
%\end{equation}
%This technique places sine-Gaussian tiles from $30-120\Hz$ above the final frequency of a given template, waveform, spaced in intervals of $15 \Hz$. It implies that high mass systems are only enters into calculation, otherwise, it returns the reweighted SNR statistic. The value of ranking statistic signifies how likely the data is to contain a signal. Therefore, we accept those triggers that lie above a pre-decided reweighted SNR threshold.

A gravitational wave signal is expected to appear in all detectors at the same time (after accounting for the light travel time between two detectors) and also expected to be generated by the same template. This argument enable us to perform a coincidence between the triggers from any two detectors. We perform the coincidence test for the arrival time and the template parameters for each trigger. The triggers that do not pass the coincidence test are discarded and the rest are passed on to the next stage of the pipeline. 

For any event reported by a search pipeline, we assign a statistical significance to express detection confidence. The search pipelines measure a false-alarm rate as a function of the detection statistic value to label the statistical significance of the detection. The $\pycbc$ search pipeline applies a time shift between the triggers generated by one detector and the other one. As the false-alarm rate calculation must be avoided from the coincidence of the pair of triggers produced by the actual gravitational wave signal, we set the minimum time shift is larger than the time window of the coincidence test. We usually choose the minimum time window of 0.1s, which is greater than the light travel time between any two detectors of the LIGO-Virgo network. Let us consider the two detectors scenario with an observation time of 10 days. We can perform the time-shift procedure 8,640,000 times, which is equivalent to a noise background time of 236,712 years. If $\hat{\rho}_\ast$ is the detection statistic value for a candidate event and there are $n_b$ noise coincident triggers above $\hat{\rho}_\ast$, then the false-alarm rate for that event is $n_b$ per 236,712 years.
%We quantify the number coincident triggers produced by the detector noise 

Since the ability to detect a gravitational wave signal depends on its amplitude and features, we measure the sensitivity of a search pipeline by generating a large number of compact binary sources assuming an astrophysical model. The sensitivity is often measured as the fraction of sources that are recovered below a given FAR value ($\mathcal{F}$). This fraction depends on the distance to the source, orientation, sky position, and masses. Therefore, we estimate a volume around a detector (or a network of detectors) in which the sources are detectable. The formula is given as,
\begin{equation}
\displaystyle \mathscr{V}(\mathcal{F}) = \int  \epsilon(\mathcal{F}; \vec{x}, \vec{\lambda}) \, p_{\rm{pop}}(\vec{x}, \vec{\lambda})   \: d\vec{x} \, d\vec{\lambda}, 
\end{equation}
where $\epsilon$ is the probability of recovering a signal produced by source of which extrinsic and intrinsic parameters are represented by $\vec{x}$ and $\vec{\lambda}$, respectively. The function $p_{\rm{pop}}(\vec{x}, \vec{\lambda})$ is the distribution function of an expected astrophysical population. If the distribution of the simulated signals matches with the expected astrophysical distribution $p_{\rm{pop}}$, then the approximate sensitivity volume is simply the injected volume $V_{\rm{inj}}$ (in which the injections are populated) rescaled by the ratio of number of found injection $N_{\rm{rec}}(\mathcal{F})$ below FAR $\mathcal{F}$ to the total number of injections ($N_{\rm{inj}}$):
\begin{equation}
\label{eq:sensitivity_vol}
\displaystyle \mathscr{V}(\mathcal{F}) \approx V_{\rm{inj}} \frac{N_{\rm{rec}}(\mathcal{F})}{N_{\rm{inj}}} 
\end{equation}
For this work, we use the weighted Monte Carlo integration method to carry out this estimation~\cite{Tiwari:2017ndi}, which is implemented in the $\pycbc$ software library. The estimation of sensitivity volume for a given population model leads to a high computational cost. The weighted Monte Carlo method can estimate the sensitive volume for multiple population models using a single set of generic injections. We carry out the sensitivity estimation assuming the sources are uniform in volume.

\section{Waveform Model and search parameter space}
\label{sec:PS and WM}
%In this section, we describe the parameter space over which we are conducting the search and waveform model we are using. As mentioned in the previous section, the parameters of the incoming gravitational waves are not know \textit{a priori} but we expect to detect them if they are within the parameter space over which we are conducting the search. For example, if we are conducting the search for IMBH blackholes we must make sure that our parameter space cover the mass range for IMBH black holes. The search makes use of a template bank which is a set of discrete points inside the parameter space. Each point in this set represents one template in the template bank. We explain in detail how these points or templates are chosen in section \ref{sec:bank}. 

In this section, we motivate and discuss the binary black hole parameter space and the waveform model that we will use in this work. We demonstrate the beyond-GR search for a specific region of the parameter space. However, this search method can be applied for any parameter space of interest.

We target the recently accomplished first and second observation runs by the network of Advanced LIGO and Advanced Virgo detectors. No GW signals were observed from the merger of intermediate-mass black hole (IMBH) binaries, which are considered to have a total mass larger than $100 \msun$~\cite{LIGOScientific:2016dsl, GWTC1-PhysRevX.9.031040}. Also, no black hole component was found with a mass less than $3 \msun$. We restrict the source population of BBH systems with a total mass between $3 \msun$ and $100 \msun$. The dimensionless spin of the black holes are uniformly distributed within the bounds imposed by the Kerr limit, $| c\vec{s}/Gm^2 | \leq 1$, where $\vec{s}$ and $m$ are the spin angular momentum and mass of the black hole, respectively. The previous studies showed that the searches of spinning BBH systems using a non-spinning template waveform is suboptimal for advanced LIGO data~\cite{Capano:2016dsf}. We incorporate the spin effects in the template waveform with a dimensionless spin range, $-0.99 \leq \chi_{1,2}\leq 0.99$, where the quantity $\chi_i$ represents the dimensionless spin of the $i\nth$ object. We encapsulate the ranges of these parameters in Table~\ref{tab:GRParamSpace}.

%\begin{table}[ht]
%    \caption{The parameter space over which the GR template bank is constructed. The GR template bank is used as a seed bank when we generate a non-GR template bank.}
%    \label{tab:GRParamSpace}
%    \begin{ruledtabular}
%    \begin{tabular*}{\columnwidth}{l@{\extracolsep{\fill}} l }
%    \hline\hline
%         \sf{Parameter}\Tstrut\Tstrut       				& \sf{Limits}  	\\ \hline
%         Compononent masses: ($m_{1,2})$/\si{\solarmass}    & $[3, \, 100]$   \\ %\hline
%         Total mass: ($m_1 + m_2$)/\si{\solarmass}   		& $[6, \, 100]$   \\ %\hline
%         Mass ratio: $q = (m_1/m_2)$ 						& $[1, \, 10]$		\\ %\hline
%         Component spins: $\chi_{1,2}$                      & $[-0.9899, \, +0.9899]$    \\ %\hline
%         Frequency band/\si{\hertz} \Bstrut\Bstrut     		& $[30, \, 1024]$    \\
%    \hline\hline
%    \end{tabular*}
%    \end{ruledtabular}
%\end{table}

\begin{table}[ht]
    \centering
   \begin{tabular}{l  c }
    \toprule[1pt]
     \toprule[1pt]
         Parameter      			\ \ \	& \ \ \ Limits  	\\
          \midrule[1pt]
         Compononent masses    & $m_{1,2} \in [3, \, 100] \msun$   \\ %\hline
         Total mass  		& $ M \in [6, \, 100] \msun$   \\ %\hline
         Mass ratio 						& $m_1/m_2 \in [1, \, 10]$		\\ %\hline
         Component spins                      & $ \chi_{1,2} \in [-0.9899, \, +0.9899]$    \\ %\hline
     %    Frequency band/\si{\hertz} \Bstrut\Bstrut     		&  $[30, \, 1024]$    \\
\bottomrule[1pt]
\bottomrule[1pt]
    \end{tabular}
    \caption{The parameter space over which the GR template bank is constructed. The GR template bank is used as a seed bank when we generate a non-GR template bank.}
    %\end{ruledtabular}
    \label{tab:GRParamSpace}
\end{table}

%The coalescence of compact binaries can be divided into three major parts -- inspiral, merger, and ringdown. In the inspiral part, the two binaries are revolving around the common center of mass while releasing energy in form of gravitational waves. In the merger part, the two binaries collide and merger to form a new compact object. In the ringdown part, the newly formed object releases some more energy and settles down. To conduct a template based search, we need to have access to the template waveform which describe the evolution of the spacetime around the binaries. The template waveforms for gravitational waves are the solutions of the Einstein field equations; however, to date there is no analytical approach available to solve the field equations so we have to resort to numerical methods or approximation. Depending on the strength of the gravitational field and system's dynamic complexity, there are three different approaches. 

%{\color{blue}{which three approaches are you talking about @Soumen can you please confirm? Post-Newtonian method, Numerical realtivity, and BH perturbation theory}}

The gravitational waves produced by the orbiting compact binaries are the solution of Einstein field equations. To date there is no analytical approach to provide the full gravitational wave solution alone. We rely on three approaches to avail the full gravitational waves, depending on the strength of gravitational fields and the system’s dynamical complexity. The inspiral and ringdown stages of a BBH coalescence are modelled using perturbative approaches: post-Newtonian expansion and quasi-normal mode expansion. The binary system develops a strong gravitational fields during the merger stage, where the accuracy of the PN expansion decreases. It's accurate description requires the NR simulation of Einstein's equation. Recent advances in perturbative approaches~\cite{Blanchet:2013haa, Berti:2005ys} and numerical relativity~\cite{Pretorius:2005gq, Campanelli:2005dd, Baker:2005vv} has made it possible to construct an accurate semi-analytical waveforms describing the entire evolution of a BBH coalescences. The recent Advanced LIGO's searches employed the semi-analytical techniques to model the waveforms: effective-one-body approach~\cite{Buonanno:1998gg, Buonanno:2009qa, Taracchini:2013rva} and phenomenological approach~\cite{Ajith:2007qp, Husa:2015iqa, Khan:2015jqa}. In this work, we employ a phenomenological waveform approximant, IMRPhenomD. This approximant is calibrated against numerical relativity simulations and it is faithful in modelling GWs from all stages of a binary black holes merger with non-precessing spins.
\begin{figure*}
\centering
    \includegraphics[width = 0.90\textwidth]{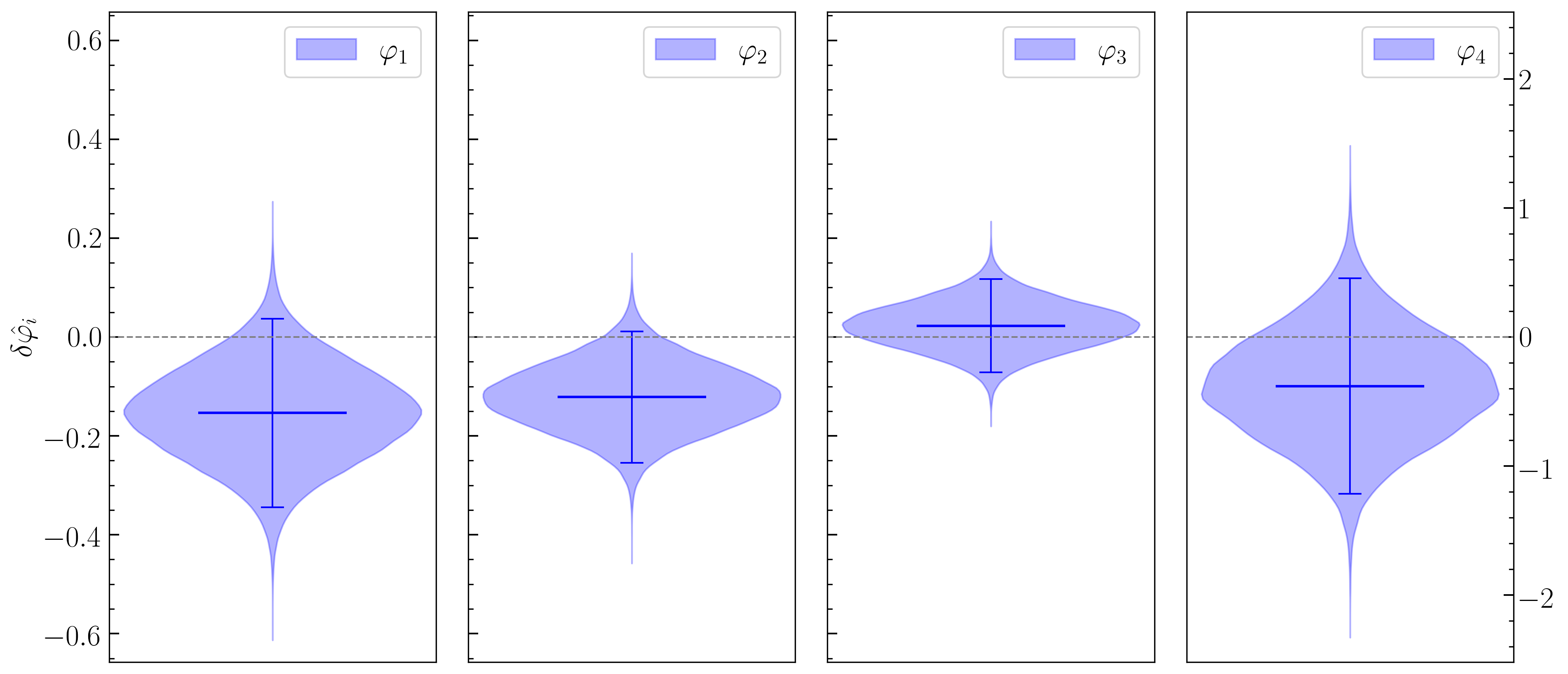}
    \caption{Combined posterior distribution for the deviation parameters $\delta\hat{\varphi}_i$. We are considering only those events which have FAR < $(1000 \ yr)^{-1}$ and exceed SNR threshold ($>6$) in inspiral regime (GW150914, GW151226, GW170104, GW170608, and GW170814). The SNR constraint is due to fact that the PN approximation is not very effective beyond inspiral regime. The dashed line corresponds to the value predicted by GR. And the vertical bar show the 90\% credible interval. This figure is reproduced by using the results from \cite{TGR-gwtc-1}.}
    \label{fig:jointDist}
\end{figure*}

Recent tests of parametrized deviation from GR by LIGO-Virgo collaborations were performed on the detected events to bound the physical effects for the coefficients that enter into the inspiral, merger, and ringdown phase~\cite{LIGOScientific:2016lio, TGR-gwtc-1}. Such test works with the TIGER framework~\cite{Li:2011cg, Agathos:2013upa, Meidam:2017dgf}, where the template waveforms for deviation from GR are modelled by introducing a fractional change in any of the GW phase coefficients of the baseline GR waveform.

If a gravitational wave signal carries a significant amount of non-GR physical effects  it could be missed by the GR template based search pipelines. Searches with accurate waveform models constructed in a top-down fashion from first principles is not possible due to a lack of knowledge of beyond-GR theories. It is therefore interesting to design a search using the parameterized waveform. In this work, we focus on the inspiral phase that is generally modeled using TaylorF2 phase $(\Phi(f))$ expansion~\cite{Poisson:1995ef, Buonanno:2009zt}:
%It is quite a challenging task, within the exact theory of general relativity, to analytically solve the two-body problem and construct the template waveforms which could be then be used to perform matched filtering. We have to resort to approximation methods. One such approximation method is the \textit{post-Newtonian} (PN) expansion, which is an expansion of the orbital quantities in the powers of a small velocity parameter $(v/c)$. The template waveform can be written in Fourier domain as $h(f)=A(f)e^{-i\Phi(f)}$, where $A(f)$ is the amplitude and $\Phi(f)$ is the phase factor. The phase factor $\Phi(f)$ can be written using PN expansion as: 
%
%\begin{widetext}
%\begin{equation}
%    \Phi(f) = 2\pi f t_c - \phi_c + \frac{3}{128\eta v^5}\left[\sum_{k=0}^7 \phi_k \left(\frac{v}{c}\right)^k + \sum_{kl = 0}^7 \phi_{kl} \left(\frac{v}{c}\right)^{kl} \text{ln} v\right]
%\end{equation}
%\end{widetext}
%
%\begin{equation}
%\begin{split}
%    \Phi(f) = &  2\pi f t_c - \phi_c + \\
%    & \frac{3}{128\eta v^5}\left[\sum_{k=0}^7 \phi_k \left(\frac{v}{c}\right)^k + \sum_{\ell = 0}^7 \phi_{\ell} \left(\frac{v}{c}\right)^{\ell} \ln{v} \right]
%    \end{split}
%\end{equation}
%
\begin{equation}
\centering
    \Phi(f) =   2\pi f t_c - \phi_c - \frac{\pi}{4} + \sum_{k=0}^7 \left[ \varphi_k +  \varphi_k^{(\ell)} \ln{f} \right]f^{\frac{k-5}{3} } 
\end{equation}
where, $t_c$ is the coalescence time, $\phi_c$ is a constant phase offset. %, $\eta \equiv m_1m_2/M^2$ is the symmetric mass ratio. , $x \equiv (\pi M f)^{1/3}$ is the instantaneous velocity.  , $M\equiv m_1+m_2$ is the total mass of the system in the detector frame, where $m_1$ and $m_2$ are the component masses. 
The quantities $\varphi_k$ and $\varphi_k^{(\ell)}$ represent the PN phase coefficients.\footnote{The index $k$, denoting the ($k/2$)-th PN order, runs \mbox{from $0$ to $7$}. %(including the terms with logarithmic terms at the 2.5PN and 3PN orders). %The value of $k$ indicates the power of $v/c$ beyond the Newtonian 0PN term.
} 
They are functions of physical parameters: masses and spins of the black holes that GR predicts. The parameterized test works by introducing a fractional deviation to a specific PN coefficient, $\varphi_k = (1 + \delta\hat{\varphi}_k) \varphi_k$. At the same time, all other coefficients remain the same as GR predicts. Under the assumption that GR is correct, the posteriors are expected to be consistent with $\delta\hat{\varphi}_k = 0$ within statistical fluctuations. On the other hand, it is likely to deviate from zero if an alternative theory is more accurate than GR. %{\hn{Blame it on LVC}} 
The parametrized test for GWTC-1 by LVC was performed on each event in the catalog to impose bound for individual cases~\cite{TGR-gwtc-1}. Finally, combined-event analysis was performed considering a list of loud events that have SNR at least 6 in the inspiral part of their signal. The posteriors of $\hat{\varphi}_k$ were combined, assuming that the parametrized violation is constant across all events. We illustrate the combined posterior distributions of $\delta\hat{\varphi}_k$ in Fig.~\ref{fig:jointDist} and report their 90\% interval in the Table~\ref{tab:nonGRparams}.\footnote{The posterior samples are taken from the public data provided by LVC~\href{https://dcc.ligo.org/LIGO-P1900087/public}{https://dcc.ligo.org/LIGO-P1900087/public}}

%Here $\Phi(f)$ denotes the phase of the gravitational wave in the frequency domain, $c$ denotes the speed of light, $t_c$ and $\phi_c$ are respectively denote the epoch and corresponding phase marking the coalescence of the two black holes,  $\eta \equiv m_1m_2/M^2$ is the symmetric mass ratio, $v \equiv (\pi M f)^{1/3}$ is the instantaneous velocity, $M\equiv m_1+m_2$ is the total mass of the system in the detector frame, where $m_1$ and $m_2$ are the component masses. $\phi_k$ are the post-Newtonian expansion coefficients. 

\begin{table}[ht]
    \centering
       %\begin{tabular*}{0.9\columnwidth}{c@{\extracolsep{\fill}} l}
    \begin{tabular}{l c }
    \toprule[1pt]
     \toprule[1pt]
         PN Deviation Parameter	\ \ 	&  \ \ Limits 			\\
         \midrule[1pt]
       \  0.5~PN 							& $ \delta\hat{\varphi}_1 \in \left[-0.345, \, 0.037\right]$ 	\\
       \ 1.0~PN 							& $\delta\hat{\varphi}_2 \in [-0.254, \, 0.011]$ \\
       \  1.5~PN 							& $ \delta\hat{\varphi}_3 \in [-0.071,	\, 0.118]$ \\
       \  2.0~PN                			& $\delta\hat{\varphi}_4 \in [-1.216,  \, 0.456]$ \\
    \bottomrule[1pt]
     \bottomrule[1pt]
    \end{tabular}
    \caption{90\% credible interval of the  deviation parameters obtained by combining their posterior distributions from the four events GW150914, GW151226, GW170104, GW170608, and GW170814. The eight-dimensional non-GR intrinsic parameter space consists of the GR parameters shown in Table-\ref{tab:GRParamSpace} and the post-Newtonian deviation parameters shown here. This table is reproduced by using the results from \cite{TGR-gwtc-1}.}
 \label{tab:nonGRparams}
\end{table}

We limit ourselves to deviations upto the 2PN order to keep the computational cost under control. The generalization to incorporate the deviation coefficients up to the 3.5PN order is straightforward. %For the non-GR  parameter space, the ranges of mass and spins are the same as those shown in table \ref{tab:GRParamSpace}. The ranges of $\delta\hat{\phi}_i$ are given in table \ref{tab:nonGRparams}. We have used the data from GWTC-1\cite{TGR-gwtc-1} catalogue to calculate these ranges. 
In this work, we consider 4 dimensional GR parameter space: ${\vec \theta}_\text{GR} \equiv \{m_1, \, m_2, \, \chi_1, \, \chi_2 \} $ and the parameter ranges are given in table \ref{tab:GRParamSpace}. On the other hand, non-GR parameter space includes extra 4 dimensions, which is a super space of GR such that ${\vec \theta}_\text{non-GR} \equiv \{m_1, \, m_2, \, \chi_1, \, \chi_2, \, \delta\hat{\varphi}_1, \, \delta\hat{\varphi}_2, \, \delta\hat{\varphi}_3, \, \delta\hat{\varphi}_4\}$.
\section{Template Bank Construction and Validation}
\label{sec:bank}

In this section, we describe the construction of two template banks that cover the GR and non-GR parameter space, respectively. We discuss the sensitivity improvement that can be achieved by including the PN deviation parameters.

The template bank ($\mathscr{T}$) is a set of filter waveforms, each of them marked as a point in the parameter space that considered for the analysis ($\mathscr{T} \equiv \{\lambda_i \} $) ~\cite{PhysRevD.44.3819, Owen:1995tm, Balasubramanian:1995bm, Owen:1998dk}. As the parameters of an incoming gravitational wave signal are not known a priori, we filter the data against each of the waveforms in the bank and obtain the corresponding signal-to-noise ratio. The discreteness of the bank leads to a loss in signal-to-noise ratio. Therefore, the obvious demand is to place the templates appropriately so that the maximum loss in signal-to-noise ratio does not exceeds a threshold. The maximum loss is determined by the minimal match ($\mmin$) of the bank,
\begin{equation}
\mmin = \min_{\lambda^{'}}\left[ \max_{ \lambda \in \mathscr{T} }\mathcal{M}\left( h(\lambda^{'}), h(\lambda)\right ) \right],
\end{equation}   
where $\lambda^{'}$ denotes any arbitrary point in the parameter space. The quantity $\mathcal{M}$ refers to the inner product between two arbitrary normalized waveforms maximized over the extrinsic parameters $(\xi)$,
\begin{equation}
\label{eq:match}
\mathcal{M}(a, b) = \max_{\xi}\, \innerprod{a}{b} .
\end{equation}
In this work, we calculate matches by maximizing over the coalescence time, phase, distance to the source and sky location.

 In GW data analysis, we usually construct the bank at $\mmin \sim 0.97$, which corresponds to a 3\% loss in signal-to-noise ratio. However, this criterion can't tell whether a template bank is faithful for detecting an arbitrary signal ($h_a$),. We usually compute a fitting factor to quantify the faithfulness of a template bank for detecting $h_a$, which is defined as the maximum match between $h_a$ and all templates in the bank~\cite{PhysRevD.52.605},
\begin{equation}
\fFactor(h_a) = \max_{\lambda \in \mathscr{T}} \mathcal{M}\left(h_a, h(\lambda) \right)
\end{equation}
 
\begin{figure}
    %\centering
    \includegraphics[width = 0.48\textwidth]{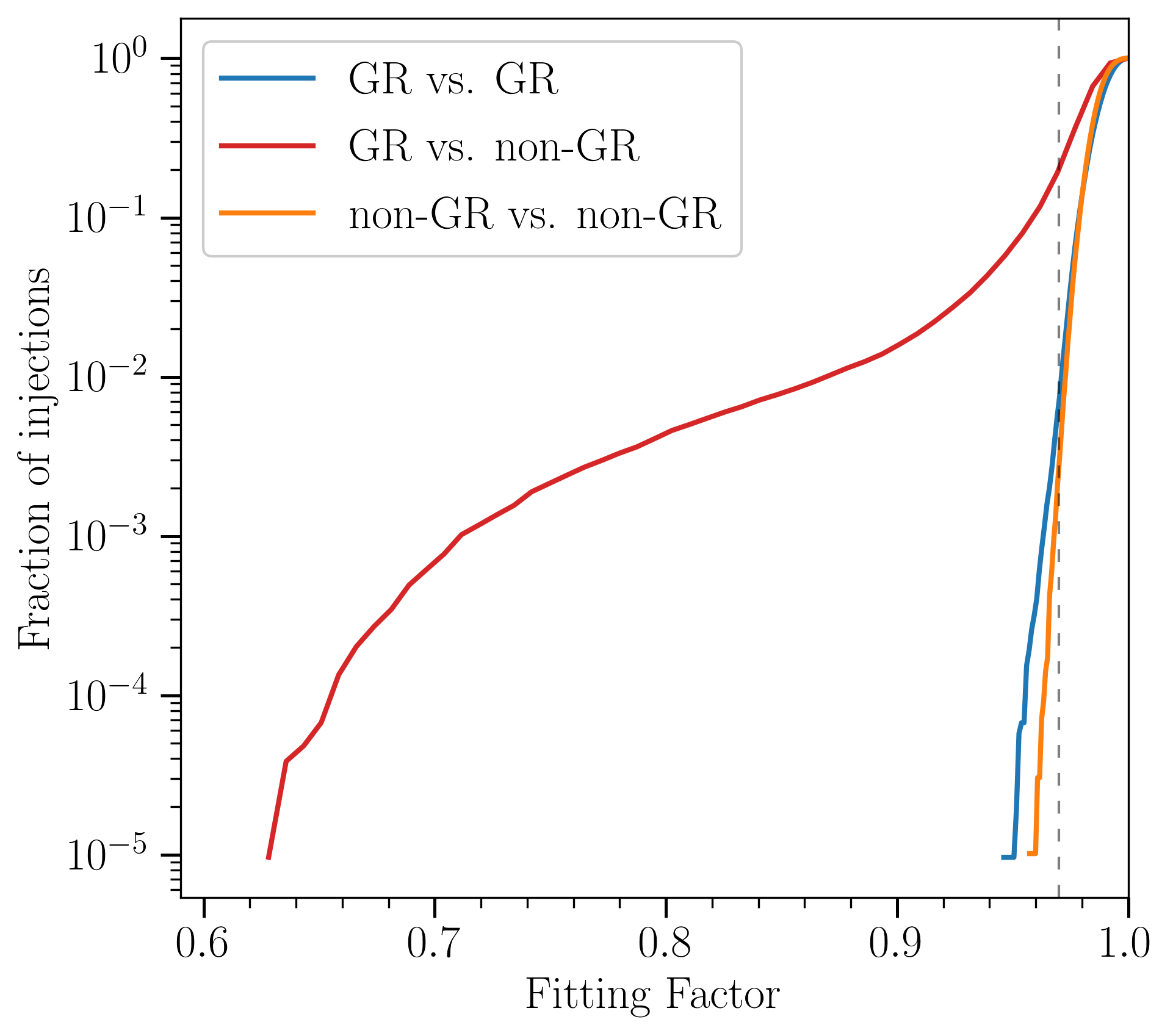}
    \caption{Cumulative distribution of fitting factors shown for three different cases: GR template bank against the GR injections (blue curve), GR template bank against the non-GR injections (red curve), non-GR template bank against the non-GR injections (orange curve). The grey dashed line corresponds to the fitting factor of 0.97 which was also the minimal match at which our banks were constructed.}
    \label{fig:fitting_factor}
\end{figure}

\begin{figure*}
    \includegraphics[width = 0.97\textwidth]{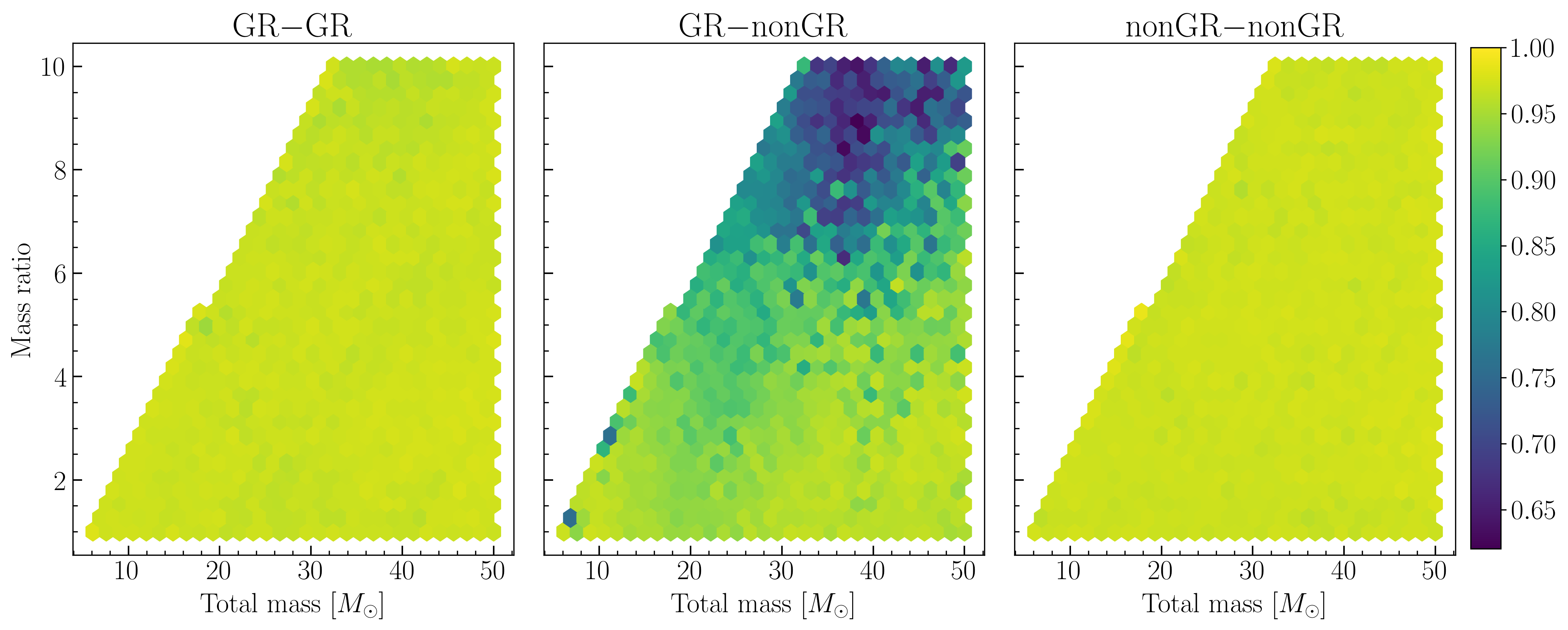}
    \caption{Fitting factor plots in total mass and mass ratio plane for three cases. The colorbar shows the \textit{lowest} fitting factor in the respective hexagonal bin. The first plot shows GR injections in the GR bank. The second plots shows GR injections in the non-GR bank. The third plot shows non-GR injections in the non-GR bank. We infer that the region around mass ratio $\geq$ 4.5 and total mass $\in [15, 50]$ in the GR bank has relatively low fitting factor against the non-GR bank (center plot) and the bank could be missing signal if they happened be in this region. }
    \label{fig:hexbinmin}
\end{figure*}
An arbitrarily denser template bank could satisfy the minimal match criteria. However, such a dense bank would lead to a computational burden and increase the expected number of false alarm candidates. The standard choice is to have an optimal template bank consisting of the smallest number of templates and adequately covering the entire parameter space. Many previous studies focused on developing the template placement algorithm. The most optimal method is an example of the sphere covering problem, which is achieved by $\anstar{n}$ lattice in flat parameter space~\cite{temp-plac-1-Prix_2007}. This approach is called a geometric placement that has been extensively used for inspiral-dominated signals, where the parameter space metric components are approximately constant~\cite{Babak:2006ty, Cokelaer:2007kx, Brown:2012qf, Harry:2013tca}. However, the metric is no more constant at all when the full inspiral-merger-ringdown waveform is considered. It has been led to the development of stochastic placement~\cite{Harry:2009ea, temp-bank-2-PhysRevD.89.084041, Privitera:2013xza}, which generates a random proposal in each iteration and accepts if it is far from existing templates in the bank; otherwise, reject it. The process continues until the rejection rate reaches a preset convergence threshold. The advantage of the stochastic method is that it can place the templates for any kind of parameter space, and the resulting template bank is robust. However, its intrinsic stochastic nature leads to an over-coverage in the resulting bank. Several recent methods have been developed to combine the space efficiency of the lattice-based geometric placement and the robustness of the stochastic approach. The straightforward way is to construct a final stochastic template bank by seeding a precomputed geometric bank of the inspiral-dominated region of parameter space~\cite{temp-bank-1, DalCanton:2017ala}. Another way is to place the $\anstar{n}$ lattice in a locally flat coordinate system where the metric components vary slowly, which is known as hybrid geometric-random template placement algorithm~\cite{hybrid-bank-1, 2018cosp...42E2899R, hybrid-bank-2}. These hybrid approaches were extensively used to analyze the data from Advanced LIGO-Virgo's first, second and third observation runs~\cite{LIGOScientific:2016vbw, GWTC1-PhysRevX.9.031040, GWTC-2-Abbott:2020niy, LIGOScientific:2021djp}. While the $\anstar{n}$ lattice is optimal for a flat unbounded parameter space, both the boundedness (from the range of search parameters) and curvature of the intrinsic parameter space affects the placement of templates. The former leads to "boundary effects" where the space efficiency of the $\anstar{n}$ lattice cannot be fully realised. On the other hand, the curvature inevitably leads to over-density due to templates that are required to cover the holes between local flat patches.

In this work, we construct the template bank for GR parameter space using the recently developed hybrid geometric-random template placement algorithm~\cite{hybrid-bank-1, 2018cosp...42E2899R, hybrid-bank-2}. This method combines the space efficiency of the $\anstar{3}$ lattice along with the robustness of stochastic method. The lattice construction is accomplished by employing a parameter space metric with a suitable coordinate system where metric component varies slowly. This method starts with a large number of random proposals uniformly sprayed over the parameter space. Thereafter, starting from a randomly chosen  point it places suitably oriented $\anstar{3}$ lattice points assuming locally flat patches within the space and remove the random proposals that lie within a distance $\dmax = \sqrt{1 - \mmin }$ of the lattice points. The placement process continues until all the random proposals are removed. For the GR parameter space with a noise PSD of Advanced LIGO's first observation run, we have found the hybrid bank consists of 27,309 templates.

%As the GR parameter space explicitly does not include the effect of PN deviation, fitting factors for the non-GR gravitational-wave signals can be lower than 0.97. 
We carry out Monte Carlo simulations to compute the fitting factors distribution against two different set of injections drawn randomly within the parameter space. First, we generate a set of GR injections assuming a uniform distribution of component masses and individual dimensionless spins of the black holes. This injection analysis would tell us whether the GR bank is adequate to cover the GR search space. Second, we include the PN deviation parameters to generate a set of non-GR injections, where the deviation parameters are chosen assuming a uniform distribution between their ranges as reported in Table~\ref{tab:nonGRparams}. As we consider the deviations in lower post-Newtonian orders and the LIGO was not sensitive at low frequencies during the first observation run, the departure for a high mass binary from the GR space is expected to be small.  We restrict the total mass of the injected binaries up to $50 \msun$. We construct the injection sets for 100,000 aligned-spin binary systems with the IMRPhenomD waveform model. 

Figure~\ref{fig:fitting_factor} shows the cumulative distribution of fitting factors for both the GR and non-GR injection sets. We can see that all the GR injections are recovered above the fitting factor of 0.95, and only 0.7\% of them are found below 0.97. This result implies that the GR template bank is effectual for detecting the signals if GR is true. However, the fitting factor curve for non-GR injections reaches 0.6, and 20\% of them are found below 0.97. This indicates the GR template bank is ineffectual for detecting the signals if they carry significant amount of non-GR physical effects. It is therefore interesting to have a non-GR template bank by incorporating the PN deviation parameters.

\begin{figure*}
%<<<<<<< HEAD
    \includegraphics[width = 0.90\textwidth]{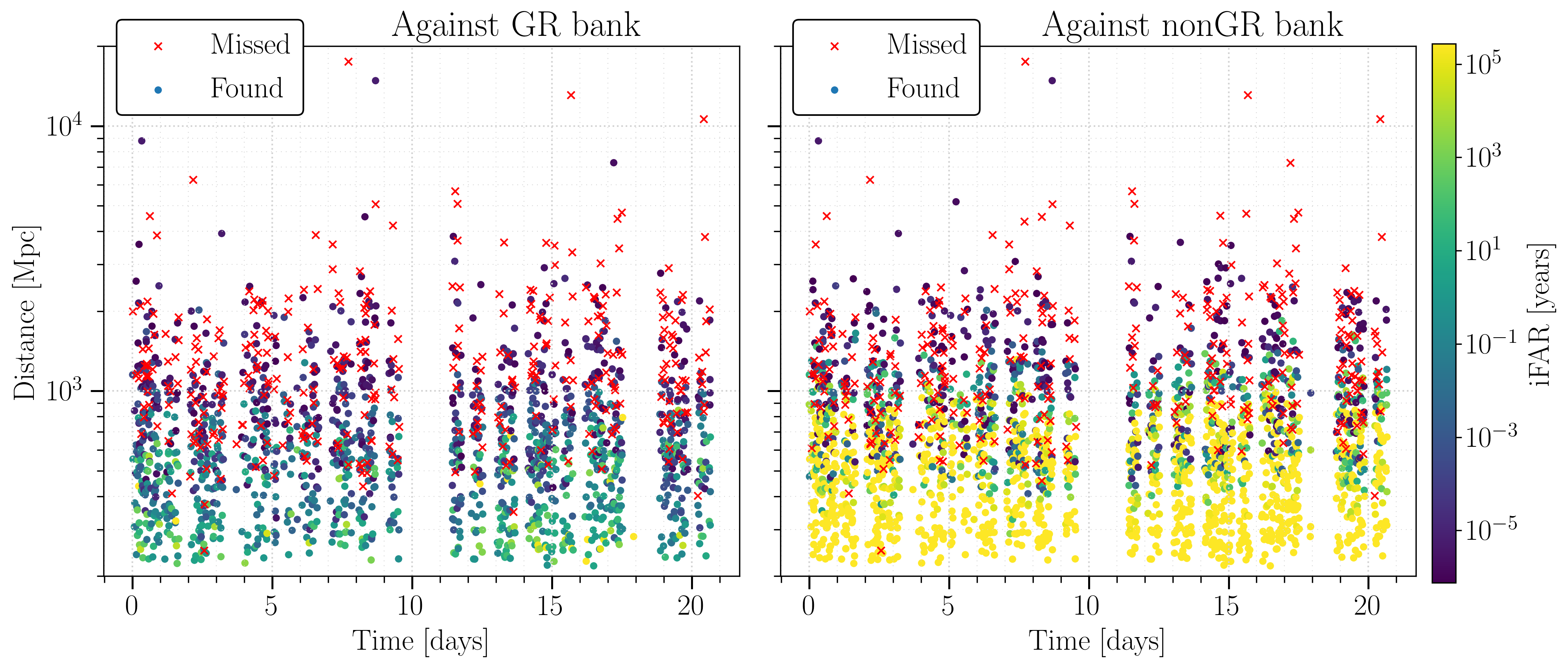}
    \caption{ The scatter plot of found and missed instances of the non-GR signal which had lowest fitting factor agains the GR bank. The cross ticks and solid circles indicate missed and found instances, respectively. The color of the circles indicates the inverse false alarm rate ($\iFAR$) of the found injections, and their values are depicted in the color bar. These results are produced using the three weeks of LIGO's O1 data starting from a GPS time of 1128812417 seconds. The vertical white bands in the triggers indicate the time of data excluded from the analysis by applying a data quality veto. }
%=======
 %   \includegraphics[width = 0.99\textwidth]{single_inj_found_missed}
  %  \caption{Simulations performed using the worst fitting non-GR injection. This injection had the lowest fitting factor against the GR bank. This injection was repeatedly injected into the detector data and then was recovered by GR and non-GR bank. The cross ticks and solid circles indicate missed and found instances, respectively. The color of the circles indicates the inverse false alarm rate (\iFAR) of the found injections, and their values are depicted in the color bar. These results are produced using the three weeks of LIGO's O1 data starting from a GPS time of 1128812417 seconds. The vertical white bands in the triggers indicate the time of data excluded from the analysis by applying a data quality veto. We infer that below ~450 Mpc all injections are recovered by non-GR bank (right) with high confidence. Such boundary is missing for the GR bank (left).}
%>>>>>>> 92631550da8ab225585c38fb2bf01258f1e4b3ad
    \label{fig:found_missed}
\end{figure*}

The construction of non-GR bank using hybrid geometric-random placement algorithm is out of scope as it requires a valid metric in 7D parameter space.~\footnote{Indeed, non-GR parameter space is 8D. As the hybrid method can provide the effectual template bank by employing effective 3D for the GR parameter space, the current version of the hybrid method requires a 7D metric for constructing the non-GR template bank.} Also, it is not known what amount of improvement can be achieved against the stochastic method using the $\anstar{7}$ lattice. For this work, we employ the flexibility of the stochastic placement method in higher dimensions to construct the non-GR template bank. We build the 8D stochastic template bank for non-GR parameter space, where the pre-computed hybrid GR template bank is a seed bank. For computing the match between a random proposal and existing templates in the bank, we use exact match function as described in Eq.~\eqref{eq:match}. We have found that the non-GR bank consists of 53,583 templates. 

As the non-GR template bank is a superset of the GR template bank, it is obvious to say that the former must be faithful for detecting the GR signals. We only consider the non-GR injection set to quantify the coverage of the bank. The solid orange curve in Fig.~\ref{fig:fitting_factor} shows the cumulative distribution of fitting factors. We can see that the minimum fitting factor value is 0.96, and 0.3\% of the injections are found below 0.97. This result indicates that the non-GR bank is effectual for detecting the signals even if they carry non-GR effects that are manifestations of deviation from GR via PN coefficients.

The cumulative distribution fitting factors does not tell us explicit details of the sensitivity of the template bank over different regions of parameter space. Therefore, it is interesting to see the fitting factor distribution over the planes of various combinations of the parameters. Fig.~\ref{fig:hexbinmin}  shows the distribution of the minimum fitting factor over the plane of total mass and mass ratio. We can clearly see that both the GR and non-GR template banks are faithful for detecting the GR and non-GR signals, respectively. However, the GR bank is not adequate to capture the non-GR signals that belong to the region with a total mass larger than $15 \msun$ and mass ratio larger than 4. We consider this boundary to evaluate the sensitivity improvement in the non-GR bank against the GR bank for detecting the non-GR signals in LIGO's O1 data.

%%%==== SECTION ====%%%
%\section{Results: Injection Runs in O1 data}
\section{Monte-Carlo simulation studies in O1 data}
\label{sec:O1 data runs}
\begin{figure*}[htbp]
    \includegraphics[width = 0.90\textwidth]{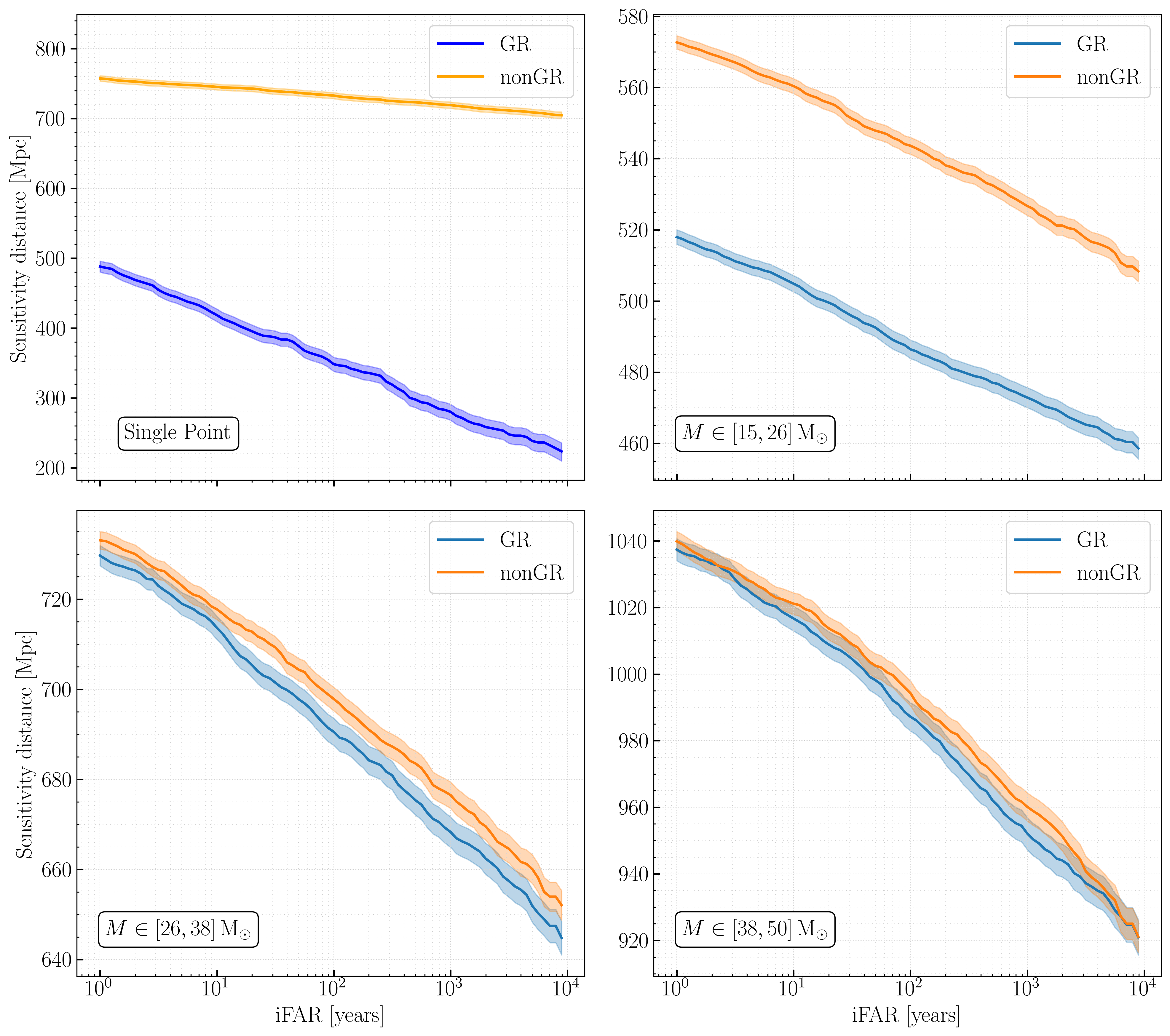}
    \caption{Comparison of sensitivity distance as a function of inverse false-alarm rate ($\iFAR$) between non-GR bank and GR bank for searching the non-GR signals. These plots were made by performing injection over a reduced range of parameters: $M \in [15, 50]$ and $q \in [4.5, 10]$. %The injection parameters and ordering of the plots is as follows. Top-left: worst fitting injection. For this type of injections the non-GR bank is able to probe twice as much distance as GR bank. Top-right, bottom-left, and bottom-right: Randomly distributed injection with mass range 15 to 26 $M_\odot$, 26 to 38 $M_\odot$, 38 to 50 $M_\odot$. For randomly distribute injection, the sensitivity of non-GR bank compared to GR-bank decreases as we go to higher mass ranges. The shaded region show ?? error bar.
    The top-left panel shows the results for worst fitting injection, where the non-GR waveform has maximum deviation from GR. For this type of injections, we can see the non-GR bank can observe two times larger distance than the GR bank. The remaining three panels are correspond to the randomly distributed injections binned in total mass. The solid line shows the median of the distribution and the shaded region shows the $\pm 1 \sigma$ width of the uncertainty. The sensitivity of non-GR bank compared to the GR-bank decreases with the total mass of the binaries.}
    \label{fig:random_injection_vt}
\end{figure*}

In the previous section, we have demonstrated the sensitivity increase when the search template waveform includes the non-GR effects, assuming the case of zero noise realization. However, the data readout from the gravitational wave detectors contains non-Gaussian non-stationary noise, which significantly fluctuates the observed signal-to-noise ratio. To consider non-Gaussian effects in evaluating the significance of events, We perform an injection study using the three weeks of advanced LIGO's O1 data surrounding the GW150914 event time starting from a GPS time of 1128812417 seconds. We exclude some part of the data from the analysis by applying a data quality veto~\footnote{The data quality veto files are available here~\href{https://www.gw-openscience.org/O1/}{https://www.gw-openscience.org/O1/}} used in O1 analysis by LVC collaboration~\cite{LIGOScientific:2016gtq, LIGOScientific:2016vbw, Nuttall:2015dqa}.  We consider two different sets of non-GR injection: i) a set of the worst fitting injection ii) a set of randomly distributed injections. We explain how the injection parameters are chosen for both runs in the following paragraphs. We also present the open box results for the 10 days of data analyzed using GR  and non-GR bank. %The stretch of of data was chosen in such a way that it covered the GW150914 event. The aim was to observe how an existing event fares against both banks. 

\begin{figure}
    \includegraphics[width = 0.48\textwidth]{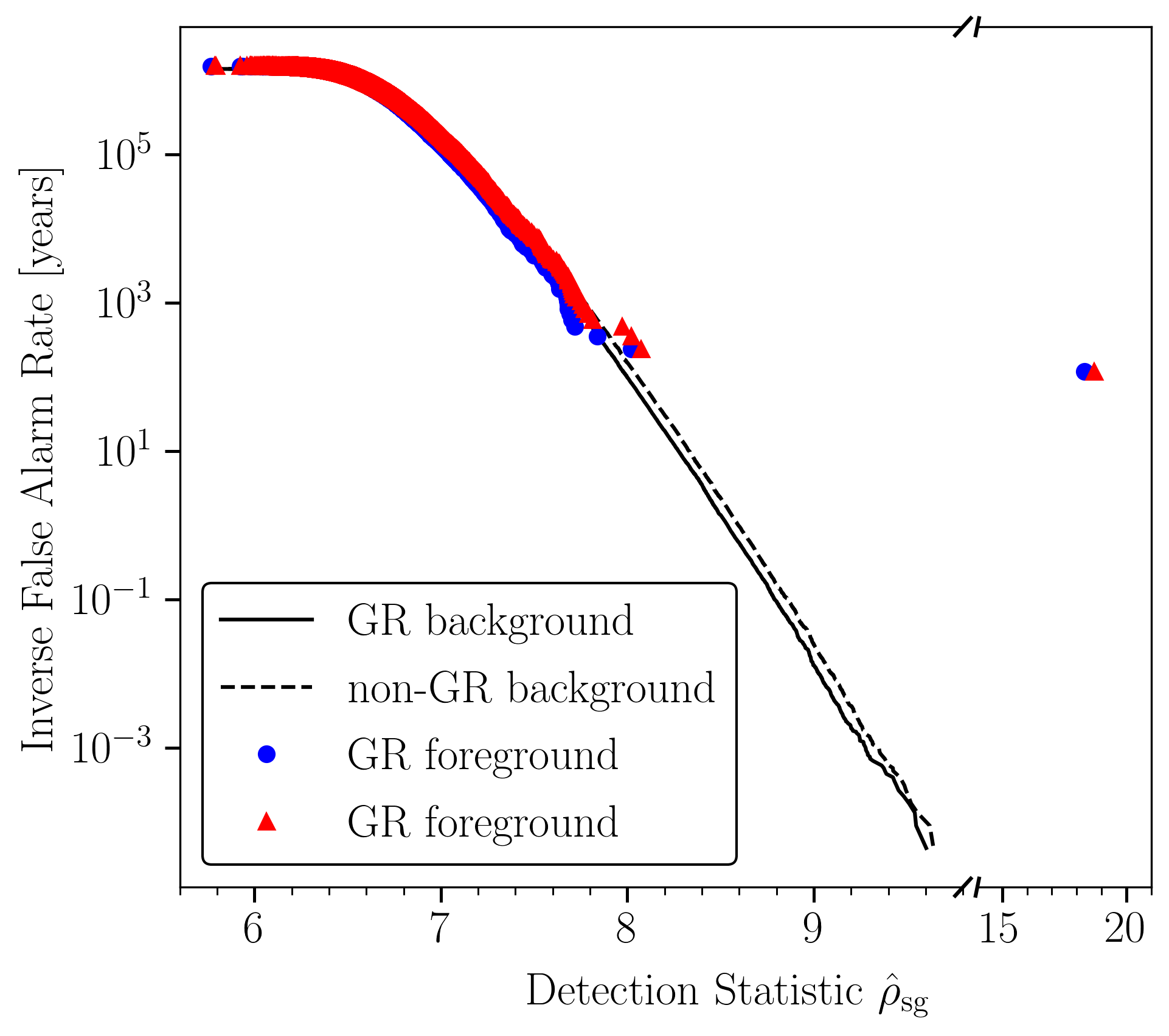}
    \caption{Comparison of search results between the non-GR bank and GR bank obtained using the ten days of O1 data surrounding the time of GW150914 event. The inverse false alarm rate is shown as a function of the detection statistic used by the search. The solid black and dashed black lines show the background of false positive detections when the search is performed using GR bank and non-GR bank, respectively. The blue and red markers show the search results obtained using GR bank and non-GR bank, respectively.    }
    \label{fig:gr_ob}
\end{figure}

%%%==== SUB-SECTION ====%%%
\subsection{The worst fitting injection}

We pick an injection from the non-GR injection set as discussed in Section~\ref{sec:bank}, which is found to have the lowest fitting factor (0.62) against the GR template bank. We repeatedly inject that non-GR waveform over three weeks of O1 data \footnote {The O1 data was obtained from Gravitational Wave Open Science Center (GWOSC)} starting from $22^{nd}$ September 2015 (GPS start time 1126989824 and GPS end time 1128804224).
%This non-GR injection is the worst fitting injection in the sense that it was recovered with the lowest fitting factor (0.62) in the exercise performed in~\ref{sec:bank}.
The intrinsic parameter values of that injection are $m_1 = 34.68 M_{\odot}, \ m_2 = 3.92 M_{\odot}, \ s_{1z} = -0.97, \ s_{2z} = -0.65, \ \delta\hat{\phi}_1 = -0.33, \ \delta\hat{\phi}_2 = -0.16, \  \delta\hat{\phi}_3 = 0.098, \ \delta\hat{\phi}_4 = -0.66$. We place the injections with a uniform time step of 256 seconds. We choose the extrinsic parameters in such a way so that the optimal SNR of the injections has uniform distribution lies between 4 and 20.

Fig.~\ref{fig:found_missed} shows the results of missed and found injections against the injected distance. 
The color of the solid circles represents inverse false alarm rate of the found injections. We have seen that  the number of found injections by non-GR bank is 1.45 times larger than the case of GR bank. The right panel of Fig.~\ref{fig:found_missed} indicates that almost all the injections below 450 Mpc are recovered with very high confidence using non-GR bank. On the other hand, such transition boundary is absent for GR bank as shown in the left panel. % for the distance range considered or may appear at a much closer distance.

To quantify how much more sensitive the non-GR bank is than the GR bank, we obtain the sensitivity volume for each search using Equation~\eqref{eq:sensitivity_vol} with the weighted Monte Carlo method. The top-left panel of Fig.~\ref{fig:random_injection_vt} shows the sensitivity distance as a function of inverse false-alarm rate ($\iFAR$). We can see the sensitive distance for the non-GR bank at an $\iFAR$ of 100 is twice as large as the GR bank. It implies that the non-GR bank can probe substantially more volume than the GR bank if we allow departures from GR.

%To quantify how much more sensitive the non-GR bank is compared to the GR bank, we report the sensitivity distance for each search as a function of iFAR in figure \ref{fig:random_injection_vt} (top-left). The sensitive distance  is a figure of merit to quantify the sensitivity of the search. For example,  iFAR of 100, the sensitive distance for non-GR bank is twice as large as the GR bank, which means that with a non-GR bank we could probe 8 times more volume compared to the GR bank - which, when multiplied with the observation time and merger rate density, translates to number of detections of the event of this particular sort. Note that this is an extreme case. We consider a more general example in the injection run performed in  \ref {sec:O1 data runs} B. {\color{red}{Soumen: Add a short description of the weighted MC and cite Vaibhav Tiwari's paper}}

\subsection{Randomly distributed signal injections}

Further, we target the potential region of the parameter space, where GR bank is inadequate to recover the non-GR signals in the sense of fitting factor as discussed in Sec.~\ref{sec:PS and WM}. We create a set of injections assuming uniform distribution over total mass, mass ratio, spins, and PN deviation parameters. We consider the total mass between $15\msun$ and $50 \msun$,  mass ratio between 4.5 and 10, the magnitude of dimensionless spin between 0 and 1, and the ranges of PN deviation parameters are described in Table~\ref{tab:nonGRparams}. We choose the extrinsic parameters in such a way so that the optimal SNR of the injections has uniform distribution lies between 4 and 20. We inject the waveforms assuming uniform time steps on 10 days of O1 data starting from 13th October (GPS start time 1128812417 and GPS end time 1129753822).

%For this injection run, we inject signals which have total mass in the range  $15 \leq M/\si{\solarmass} \leq 50 $ and mass-ratio between $4.5 \leq q \leq 10$. We are using the same stretch of data. We choose this boundaries for $M$ and $q$ since this region the GR bank has relatively low fitting factor (see figure \ref{fig:hexbinmin} (center)) compared to the non-GR bank  (see figure \ref{fig:hexbinmin} (right)). Through this injection campaign, we want to test how they perform in presence of detector noise. The injections are uniformly distributed over total mass and mass ratio. The component spins are uniformly distributed from 0 to 1 and the optimal SNR is uniformly distributed between 4 to 20.

 We present the results of this run in figure \ref{fig:random_injection_vt}. We again plot the sensitive distance as a function of the $\iFAR$ rate. From Figure~\ref{fig:random_injection_vt}, we infer that the effect of deviation parameter is more prominent in low total mass range (top right) compared to the high total mass range (bottom row). For example, at a fixed $\iFAR$ of 100, the non-GR bank is able to access $\approx 12\%$ more distance then the GR bank, which would translate to $\approx 40\%$ more volume. This number directly translates to the number of detections when we fold in the merger rate density and the observation time. For the high total mass bins $26\leq M\leq 50$, both banks perform roughly the same. This is consistence with what we saw in the fitting factor plots (see figure \ref{fig:fitting_factor}). As we go towards higher mass systems, the effect of deviation parameter vanishes.    
%1.24 = (8952/7233)

%%%==== SUB-SECTION ====%%%
%\subsection{Open box results}
%\section{Analyzing ten days of O1 data}
\section{Analysis of O1 data}
\label{sec:O1_analysis}
%\begin{figure}
 %   \includegraphics[width = 0.49\textwidth]{figures/Comparison_waveform-6.png}
  %  %\includegraphics[width = 0.90\textwidth]{figures/Comparison_waveform-6.png}
  %  \caption{This plot shows the best fitting GR and non-GR waveform in the time-domain. The intrinsic parameters of both these waveforms are different and they are mentioned in the table \ref{tab:loudest_events_GR_nonGR}. The non-GR waveform detected the GW150914 event with marginally higher SNR than the GR waveform. We also report that the match between the two template waveforms is 0.96. \hn{@Soumen can I ask you to remake this figure? I am not able to get Times New Roman font in the cluster, silly problems. The label and legend look different from other plots}}
  %  \label{fig:overlap}
%\end{figure}

We present open box results for data analysed with GR and non-GR banks with 10 days of data starting from 13th September, 2015. This stretch of data covers the first GW event, GW150914. The goal is to see how the existing event fares against the non-GR bank. We report that the GW150914 event is detected with slightly higher SNR by the non-GR bank. GW150914 was roughly an equal mass binary system, so we do expect both banks to detect it with similar SNR since for equal mass binaries the non-GR effect is less prominent. We also report that the best fitting template for GW150914 in the non-GR bank has non-zero deviation parameters and the match between the best fitting GR and non-GR template is 0.956. As a sanity check, we also verified that the deviation parameters ($\delta\phi_i$) of the best fitting template of GW150914 lie within the 90\% ranges for the same event reported in \cite{, LIGOScientific:2016lio, TGR-gwtc-1}. %In figure \ref{fig:overlap}, we have plotted the best fitting GR and non-GR template in the time-domain. 
The top 10 events reported by the non-GR bank are presented in descending order of their significance ($\iFAR$) in Table~\ref{tab:loudest_events_GR_nonGR}. 

The second most significant event is the same for both GR and the non-GR bank searches with identical template parameters, but reported at a higher significance in the GR search. The deviation parameters of the best matching template for this event are identically zero in the non-GR search. 

The 19th most significant event in the GR-bank search is ranked 3rd in the non-GR search, with a template having non-zero deviations at a higher significance by a factor of $6.7$.
In addition,  we report that there are candidate events (ranked 4, 7, 10) with non-zero deviation parameters  reported by non-GR bank which are completely missed in the GR search. However, they are of little significance as their $\iFAR$ values are very low. 

From Figure~\ref{fig:gr_ob}, we see that the non-GR background is slightly higher than the GR background. This is expected due to the larger degrees of freedom of the non-GR templates owing to the four additional deviation parameters, as a result of which they can match noise artefacts better as compared to the GR templates. One can also see that the increase in the background rate is marginal even though we have nearly twice as many templates in the non-GR search. This can be explained by a strong correlation in the PN deviation parameters.
%speculate that it is because we have 4 more parameters in the non-GR template also fits the glitch better than the GR-template. 

%==== O1 events table
%\onecolumngrid
\begin{table*}
    %\label{tab:ob_table}
    \label{tab:loudest_events_GR_nonGR}
    
    %\bgroup
    \def\arraystretch{1.50}%  1 is the default, change whatever you need
    
    \begin{tabular*}{\linewidth}{
    l@{\extracolsep{\fill}} 
    | S[table-format=2] 
    S[table-format=2.1] 
    S[table-format=2.2] 
    S[table-format=2.2] 
    S[table-format=2.2]
    S[table-format=2.2] 
    S[table-format=2.2]
    S[table-number-alignment = right]
    S[table-format=2.2] 
    S[table-format=2.2]
    | S[table-format=4] 
    S[table-format=2.2] 
    S[table-format=2.2] 
    S[table-format=3.2]    
    S[table-format=2.2] 
    S[table-number-alignment = right]
    }
    \hline \hline
    %-- Top row
    \multicolumn{1}{l|}{Event} 
        & \multicolumn{10}{c|}{non-GR bank search } 
        & \multicolumn{6}{c}{GR bank search} \\
        
    \cline{2-11} \cline{12-17}
    
    GPS Time 
        & \#\footnote{Event rank \#}
        & $M [\msun]$ 
        & $q$
        & \sf{$\chi_{\rm{eff}}$}
        & \sf{$\delta \hat\phi_1$}
        & \sf{$\delta \hat\phi_2$}
        & \sf{$\delta \hat\phi_3$}
        & \sf{$\delta \hat\phi_4$}
        & $\hat\rho_{\rm{sg}}$ 
        & $\iFAR$
        & \#
        & \sf{$M [\msun]$} 
        & \sf{$\ \ \ q$}
        & \sf{$\chi_{\rm{eff}}$}
        & $\hat\rho_{\rm{sg}}$ 
        & $\iFAR$ \\
        
    \hline

    % -- event 1
    \num{259 462.427}\footnote{GW150914: the first binary black hole merger event observed in advanced-LIGO detectors.}
        & 1
        & \num{50.97}
        & \num{1.37}
        & \small{\num{-0.34}}
        & \small{-0.26} & \small{-0.23} & \small{0.09} & \small{-1.17} 
        & \num{18.97}
        & \num{9.33e3} %$1.05\times 10^4$
        & 1
        & \num{68.32}
        & \num{1.90}
        & \small{\num{-0.17}}
        & \num{18.32}
        & \num{5.89e3} \\ \hline %$5.89\times 10^3$ 
        
    % -- event 2
    \num{798 299.477}
        & 2
        & 7.21
        & 1.33
        & \small{0.32}
        & \small{0} & \small{0} & \small{0} & \small{0}
        & 8.03
        & \num{8.36e-3} %$1.51\times 10^{-2}$ 
        & 2
        & 7.21
        & 1.33
        & \small{0.32}
        & 8.02 
        & \num{2.07e-2} \\ \hline %$7.18\times10^{-4}$   
    
    % -- event 3
    \num{796 951.201}
        & 3
        & 8.53
        & 1.51
        & \small{-0.83}
        & \small{-0.19}	& \small{-0.24}	& \small{0.10} & \small{0.05}
        & 7.98
        & \num{4.80e-3}      
        & 19
        & 9.20
        & 2.07
        & \small{-0.99}
        & 7.61
        & \num{7.18e-4} \\ \hline
       
    % -- event 4
    \num{819 769.188}
        & 4
        & 55.55
        & 8.85
        & \small{-0.88}
        & \small{-0.23}	& \small{-0.24}	& \small{0.10} & \small{-0.95}
        & 7.94
        & \num{3.20e-3}      
        & 
        & 
        & 
        & 
        & 
        & \num{} \\ \hline
    
    % -- event 5
    \num{663 132.169}
        & 5
        & 20.83
        & 5.87
        & \small{-0.98}
        & \small{0.01}	& \small{-0.16}	& \small{0.03} & \small{-0.87}
        & 7.92
        & \num{2.52e-3}      
        & \small{2685}
        & 20.21
        & 4.63
        & \small{-0.80}
        & 6.84
        & \num{4.26e-6} \\ \hline 
    
    % -- event 6
    \num{871 612.668}
        & 6	
        & 99.62	
        & 10.00	
        & \small{-0.99}	
        & \small{0}	& \small{0}	& \small{0}	& \small{0}	
        & 7.91	
        & \num{2.23e-3}	
        & 3	
        & 99.62	
        & 10.00	
        & \small{-0.99}	
        & 7.84	
        & \num{4.62e-3} \\ \hline
    
    % --event 7 
    \num{314 685.465}
        & 7	
        & 42.17	
        & 5.89	
        & \small{-0.91}	
        & \small{-0.32}	& \small{-0.22}	& \small{0.11}	& \small{-1.15}	
        & 7.92	
        & \num{2.07e-3}
        & 
        &
        &
        & \\ \hline
    
    % --event 8 
    \num{560 867.533}
        & 8	
        & 49.79	
        & 9.73	
        & \small{-0.98}	
        & \small{-0.26}	& \small{-0.24}	& \small{0.1} & \small{-1.1} 
        & 7.88	
        & \num{1.72e-3}	
        & \small{1000}	
        & 86.05	
        & 2.01	
        & \small{0.05}
        & 7.03	
        & \num{1.31e-5} \\ \hline
        
    % --event 9 
    \num{936 922.760}	
        & 9	
        & 20.34	
        & 5.32	
        & \small{0.53}	
        & \small{-0.3}	& \small{-0.09}	& \small{0.01}	& \small{-1.2}	
        & 7.86	
        & \num{1.34e-3}	
        & 12	
        & 21.95	
        & 5.77	
        & \small{0.39}	
        & 7.66	
        & \num{1.08e-3}
    \\ \hline
    
    % --event 10 
    \num{404 625.826}	
        & 10	
        & 69.67	
        & 9.92	
        & \small{-0.95}	
        & \small{0}	& \small{0}	& \small{0}	& \small{0}	
        & 7.86
        & \num{1.33e-3}	
        & 
        & 	
        & 	
        & 	
        & 	
        & 
    \\
    \hline\hline
    \end{tabular*}
    %\egroup
    \caption{A comparison of the top 10 events reported by the standard GR and non-GR searches, sorted by the event rank in the non-GR bank search. The GPS time of each event is reported in seconds relative to a fiducial epoch corresponding to GPS time 1126000000 \si{\second}. The events' statistical significance with respect to the background of accidental coincidences is measured by its corresponding inverse false-alarm rate ($\iFAR$) in units of \si{\year}. We have reported the reweighted-SNR detection statistic $\hat\rho_{\rm{sg}}$ as described in Sec.~\ref{sec:pipline}.}
\end{table*}
%\twocolumngrid
\section{Discussion and Future Work}
\label{sec:disc}

We present the outline of a template-based search for gravitational-wave signals from binary black holes going beyond general relativity. The non-GR template waveforms used in this search are constructed by allowing deviations of the GW phasing coefficients at different PN orders from their numerical values in GR. The extent of the deviation in the PN coefficients is per the 90\% credible interval of their posterior probability distribution, as measured from events in the GWTC-1 catalogue. Fitting factor studies show that the GR template bank could be missing such exotic signals in a region of the parameter space characterized by low total mass and high mass ratio region. This is evident from the low fitting factor values seen in  (Figure \ref{fig:fitting_factor}) that drop as low as $0.6$ in that region. With this simulation, we find that the GR bank has 20\% non-GR injections below the target fitting factor of the bank ($0.97$), which drops to a mere $0.7$\% when a non-GR bank is used (thus alluding to better coverage of non-GR signals). We replicated the same result by performing signal injections in the archival detector data from advanced-LIGO's first observing run O1. For this simulation, we report the search's sensitive distance (in Mpc) results. When a signal having the worst fitting factor against the GR bank is injected repeatedly into data, the non-GR bank is sensitive (at a fixed false alarm rate) out to a distance that is twice as far (Figure \ref{fig:random_injection_vt}) in comparison to the GR bank. While for a more diverse set of injections (Figure \ref{fig:random_injection_vt} top right), the non-GR bank has 12\% more distance reach over the GR bank, which translates to $40\%$ more sources! The improvement in the distance reach of an explicit non-GR search diminishes as we increase the total mass of the (figure \ref{fig:random_injection_vt} bottom row). This saturation is possibly due to the smaller time-frequency volume of such target signals, which have much fewer cycles in band and are well-matched by shorter templates in the bank. Note that this implies that the saturation in improving the sensitive distance for sources above a certain total mass also depends on the detector sensitivity and effective bandwidth and the astrophysical abundance (distribution) of sources. 

It should be noted that non-GR signals with a short time-frequency volume would be efficiently detected by ``burst'' search pipelines that are optimized to detect narrow band energy transients. On the other hand, non-GR signals that have more in-band cycles can be more efficiently detected by a template-based non-GR search. In this sense, our method could complement the burst searches in detecting exotic GW signals which would otherwise pass undetected, as shown by simulations in this paper. At a detector sensitivity (characterized by the O1 PSD curve used in our analysis), we find little advantage of a template-based search for non-GR signals from compact binaries having a total mass larger than $50\, \msun$. One wonders if a burst search could cover this cross-over region of the parameter space more efficiently? 

We have also presented the results from a re-analysis of 10 days of O1 data that includes the epoch of the historical GW150914 event with both GR and non-GR template waveforms. The foreground and background events from both these searches can be seen in Figure \ref{fig:gr_ob}. We expect a larger background rate for the non-GR search owing to the extra degrees of freedom of the non-GR template waveforms. From the plot mentioned above, we find that while the background rate for the non-GR search is indeed larger than the GR search, it is not phenomenally larger in comparison, even though the template bank size is almost two times bigger. This is understood from the fact that the number of statistically independent templates used in the non-GR search is only marginally larger than those in the GR bank. In other words, while the total number of templates is much larger, the effective number of independent degrees of freedom has only marginally increased in going from GR to the non-GR search. This hints at using an effective linear combination of deviation parameters to handle the high computational cost of the non-GR searches. 

A comparison of the top 10 events from both the GR and non-GR searches are tabulated in Table \ref{tab:loudest_events_GR_nonGR}, sorted by the detection statistic of the events as recorded in the non-GR search. While both the searches report the GW150914 event unambiguously as the most significant event, the detection statistic is marginally higher in the non-GR search as recorded by the ``detection template'' that has significant deviation parameters. Also, the inverse false-alarm rate ($\iFAR$) of this event bearing identical GPS time-stamps is $\sim \times 1.6$ higher in the non-GR search. As the $\iFAR$ rate measures the statistical confidence of an event's detection, we can see that the non-GR search detects the GW150914 marginally better on both counts. This observation leads us to believe that there may be some merit in the re-analysis of archival GW data using the non-GR search method outlined in this paper. Further, motivated by the stronger detection of GW150914 by a template with non-zero deviation parameters, it would also be interesting to perform a parameter estimation study of the GW150914 events by sampling deviation parameters (either separately or in a suitable linear combination). 

It is possible that physical effects not considered in the phenomenological IMRPhenomD  template waveform model (such as precession, eccentricity, higher-modes etc.) could be substantially correlated with non-GR deviation parameters. This needs to be investigated in future. 

In this exercise, we have allowed PN deviation parameters only up to 2PN order to control the computational cost. We believe that it is worth investigating the effects of higher-order PN deviation parameters since the higher-order deviation parameters ($\geq 2$ PN) seem to have an order of magnitude larger error bars than lower-order deviation parameters. The latter can be interpreted as more leeway to incorporate signals that deviate from standard GR phasing. To decrease the computational cost when higher-order PN terms are included, we could consider varying a linear combination of non-GR parameters dictated by a singular value decomposition on the Fisher information matrix over the PN deviation parameters. This would help reduce the effective dimensionality of the search parameter space and reduce the overall computational cost by reducing the non-GR template bank size. Finally, the results presented here originate from the O1 power spectral density curve used in our analysis and the range of deviations as measured from GWTC-1 events. We intend to repeat the analysis with the deviation ranges reported in the GWTC-3 since they become smaller with improved sensitivity and more number of observations. It would help us understand how robust the template based search pipeline is. %It would be fascinating to follow up on high mass ratio candidates in GW catalogues previously detected by the standard GR search by a non-GR search.

%%%==== ACKNOWLEDGEMENTS ====%%%

\begin{acknowledgements}

We acknowledge the efforts of Mohini Rangwala (Masters student from SVNIT, Surat, India) for some exploratory studies of the fitting factor between GR and non-GR signals as part of her MSc thesis. We are highly grateful for the suggestions and support received from Ian W. Harry, Thomas Dent and Stuart Anderson with running the modified PyCBC pipeline on the LIGO-Caltech cluster. We also thank several other LSC colleagues: P. Ajith, Sashvath Kapadia, K. G. Arun, Md. Saleem, K Haris, Tjonnie Li, and Chris van den Broeck for feedback and suggestions which has helped us considerably improve the scope and content of this paper. H.~N. and S. ~R. are supported by the research program of the Netherlands Organisation for Scientific Research (NWO). A.~S. gratefully acknowledges the funding received from the Department of Science and Technology, India for generous funding via it's DST-ICPS (Data Sciene) cluster project. 

We gratefully acknowledge computational resources provided by the LIGO Laboratory and supported by the NSF Grants No.~PHY-0757058 and No.~PHY-0823459. This research has made use of data, software and/or web tools obtained from the Gravitational Wave Open Science Center, a service of LIGO Laboratory~\cite{GWOSC:catalog}, the LIGO Scientific Collaboration and the Virgo Collaboration. LIGO Laboratory and Advanced LIGO are funded by the United States National Science Foundation (NSF) as well as the Science and Technology Facilities Council (STFC) of the United Kingdom, the Max-Planck-Society (MPS), and the State of Niedersachsen/Germany for support of the construction of Advanced LIGO and construction and operation of the GEO600 detector. Additional support for Advanced LIGO was provided by the Australian Research Council. Virgo is funded, through the European Gravitational Observatory (EGO), by the French Centre National de Recherche Scientifique (CNRS), the Italian Istituto Nazionale di Fisica Nucleare (INFN) and the Dutch Nikhef, with contributions by institutions from Belgium, Germany, Greece, Hungary, Ireland, Japan, Monaco, Poland, Portugal, Spain.

\end{acknowledgements}

%%%==== BIBLIOGRAPHY ====%%%
\bibliography{reference}

\end{document}